\documentclass[reprint,amsmath,amssymb,prl,]{revtex4-1}
\usepackage{graphicx}
\usepackage{dcolumn}
\usepackage{bm}
\usepackage{wasysym}
\usepackage{MnSymbol}
\usepackage[dvipsnames]{xcolor}

\begin{document}

\preprint{APS/123-QED}

\title{Giant osmotic pressure in the forced wetting of hydrophobic nanopores}

\author{Mill\'an Michelin-Jamois}
\affiliation {MATEIS, INSA-Lyon, CNRS UMR 5510, 69621 Villeurbanne , France}
\author{Cyril Picard}%
 \email{cyril.picard@ujf-grenoble.fr}
 \affiliation{%
Univ. Grenoble Alpes, LIPHY, F-38000 Grenoble, France 
}
\author{G\'erard Vigier}
\affiliation {MATEIS, INSA-Lyon, CNRS UMR 5510, 69621 Villeurbanne , France}
 \author{Elisabeth Charlaix}
\affiliation{%
Univ. Grenoble Alpes, LIPHY, F-38000 Grenoble, France 
}%

\date{\today}

\begin{abstract}

The forced intrusion of water in hydrophobic nanoporous pulverulent material is of  interest for quick storage of energy. With nanometric pores the energy  storage capacity is controlled  by interfacial phenomena. With subnanometric pores, we demonstrate that a breakdown occurs with the emergence of molecular exclusion as a leading contribution. This bulk exclusion effect leads to an osmotic contribution to the pressure that can reach levels never previously sustained.
We illustrate on various electrolytes and different microporous materials, that a simple osmotic pressure law 
accounts   quantitatively for  the enhancement of the intrusion and extrusion pressures governing the forced wetting and spontaneous drying of the nanopores. Using electrolyte solutions,  energy storage and power capacities can be widely enhanced.

\end{abstract}

\maketitle

Water in hydrophobic confinement is a topic of major importance for industrial applications, such as boiling,  heat or mass transfers at interfaces  \cite{Chen2009,Zhang2014,Eroshenko2001a,  Suciu2003}, but also for the  general understanding of hydrophobic interactions mediated by water in biological matter  \cite{Berne2009,TenWolde2002,zhou2014} and biomolecular responses under osmotic stress \cite{Masson2013}. Ordered hydrophobic nanoporous materials have been used to study water confined between hydrophobic surfaces in well defined geometries. 
But these material are also considered for novel energy applications.
Due to the large specific area of nanoporous materials, of the order of  1000 m$^{2}$/cm$^{-3}$, and to the high  tension of  liquid/hydrophobic solid interfaces, forcing  the intrusion of water  in these materials provides a way to store energy in the form of interfacial free energy
\cite{Eroshenko2001a}. 
Then, the spontaneous extrusion of water out of the nanopores, related to their drying transition, allows the partial or full recovery of the stored energy. 

In nanoporous materials of  pore diameter larger than typically  ten water molecules, it was shown that macroscopic concepts describe  quantitatively  the pressure $P_{int}$ at which the forced wetting, or intrusion, occurs. 
For instance in the cylindrical pores of hydrophobic mesoporous silicas,   the intrusion pressure obeys  the Laplace law of classical capillarity  
and scales as the inverse of the pore radius $1/R_p$ \cite{Lefevre2004a}.
The drying transition on the other hand was shown  to be triggered by  the nucleation of nanobubbles, a mechanism also governed by interfacial phenomena \cite{Leung2003,Sharma2012,Guillemot2012,Grosu2014}. This nucleation process, in confinement, has also proven to be of interest to study the growth of nanobubbles, of precisely controlled shape, in relation with line tension effects.

With subnanometric pores, such as the pores of zeolithes or Metal Organic Frameworks, a breakdown occurs \cite{Tzanis2014}. In particular, electrolyte solutions exhibit huge wetting  and drying pressures as compared to the ones of pure water.   In this letter, we demonstrate for the first time that exclusion effects emerge as a leading contribution responsible for this giant pressure increase. 
We find that during the intrusion/extrusion processes, the pressure in the bulk electrolyte corresponds to the simple addition  of the osmotic pressure $\Pi$  to the value of the intrusion/extrusion pressures obtained for  pure water. This simple osmotic effect can increase very significantly the density of the stored energy.
  
\begin{figure}[h]
\begin{center}
\includegraphics[width=\linewidth]{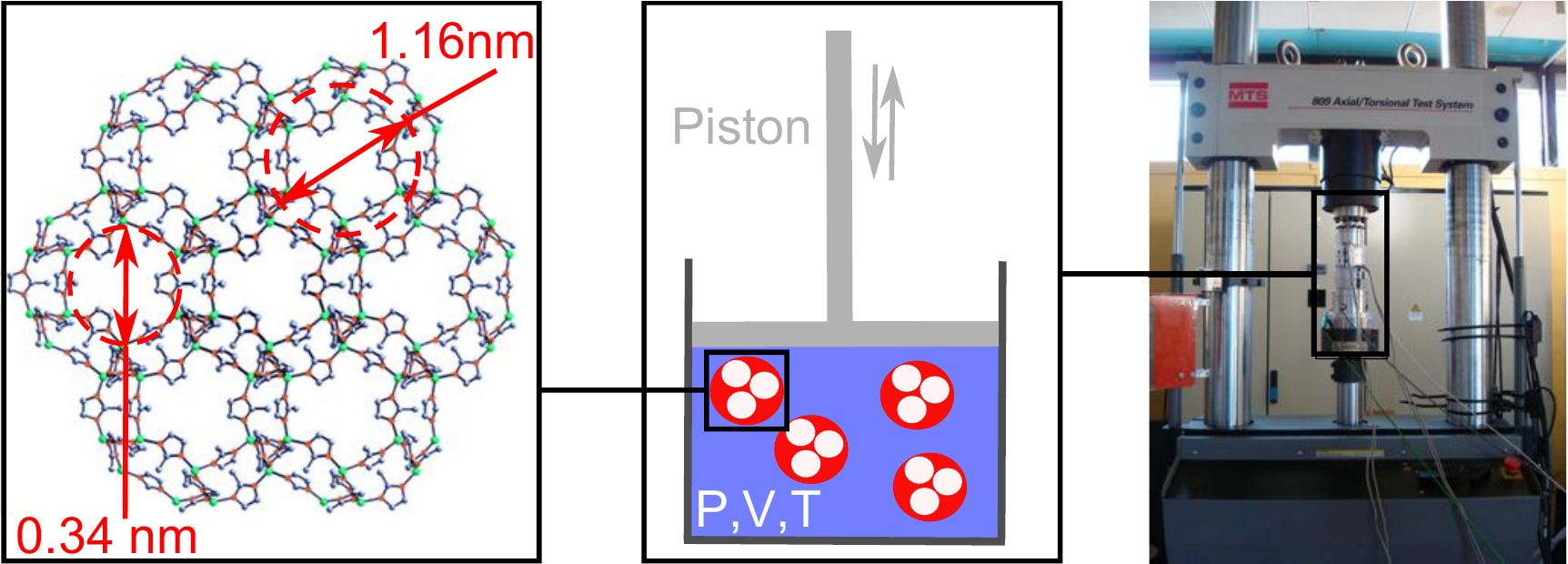}
\caption{Principle of forced intrusion/extrusion cycles. Pressure measurements are performed in an instrumented water-proof chamber, containing  ZIF-8(red)\cite{Park2006}  and electrolytes solution (blue), mounted on a traction machine.}\label{process}
\end{center}
\end{figure}

Experiments have been carried out on a Metal Organic Framework (MOF) called Zeolitic Imidazolate Framework 8 (ZIF-8)\cite{Park2006}, combined with seven different electrolyte solutions in various concentrations and temperature. ZIF-8 is commonly studied in the field of gas separation because of its high selectivity \cite{Yang2012}. 
 An important feature of ZIF-8 is its natural hydrophobicity.  Forced intrusion/extrusion cycles were achieved using a piston-cylinder device as described by Guillemot et al  \cite{Guillemot2012a} (see figure~\ref{process}), mounted on a traction/compression machine. Pressure-Volume (P-V) cycles  were obtained on previously evacuated samples at constant temperature and under quasi-static conditions. 
 
\begin{figure}[hb]
\begin{center}
 \includegraphics[width=0.9\linewidth]{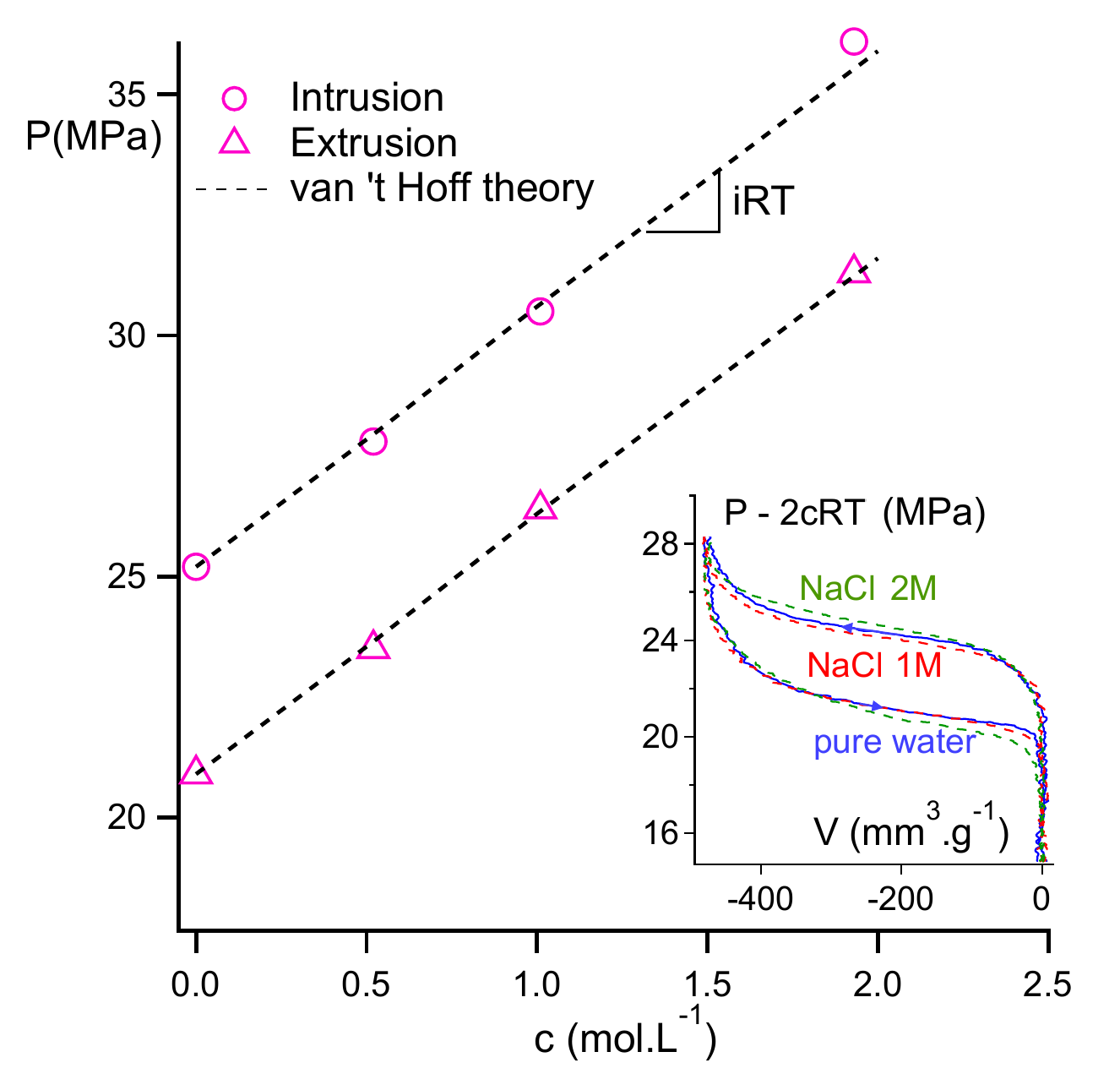}
\caption{Values of the intrusion and extrusion pressures in ZIF-8 as a function of the concentration $c$ of NaCl solutions at 323 K. The inset shows the difference between the pressure and the van 't Hoff osmotic pressure versus intruded volume for ZIF-8/water and two different ZIF-8/NaCl solutions at $T=343$ K.}
\label{NaCl}
\end{center}
\end{figure}

The pressure of the liquid versus the intruded volume for a ZIF-8/water system is presented in the inset of figure~\ref{NaCl}. Intrusion and extrusion processes correspond to plateaus of the cycle. The value of the intrusion (resp. extrusion) pressure is  defined as the average value on the intrusion (resp. extrusion) plateau. The ZIF-8/pure water system shows a small hysteresis with flat plateaus for both intrusion and extrusion. The intruded volume is about $500$~mm$^3$.g$^{-1}$. 
Values of the intrusion and extrusion pressures are respectively $25.2 \pm 0.2$~MPa and $20.9 \pm 0.2$~MPa for this pure water system. The cycle obtained for ZIF8/NaCl solutions  reproduce perfectly the pure water cycle with a simple shift in pressure (see inset of figure \ref{NaCl}). 
 As shown in figure~\ref{NaCl}, the values of the intrusion pressure $P_{int}$ and the extrusion pressure $P_{ext}$ increase linearly with the NaCl concentration $c$, with slopes of  respectively $5.5 \pm 0.2$~MPa.M$^{-1}$ for intrusion and $5.2\pm0.2$~MPa.M$^{-1}$ for extrusion. 
 This linear dependency corresponds very closely  to  the van~'t Hoff law of osmotic pressure $\Pi$:
 \begin{equation}
\Pi=icRT
\end{equation}
 where  $R$ is the perfect gas constant, $T$ the temperature and $i$ the number of ions per electrolyte molecule. The corresponding slope is $iRT=5.37$~MPa.M$^{-1}$ for $i=2$ and $T=323K$. This theoretical value is  very close to the slopes of the experimental data.
 
\begin{figure}[h]
\begin{center}
 \includegraphics[width=0.6\linewidth]{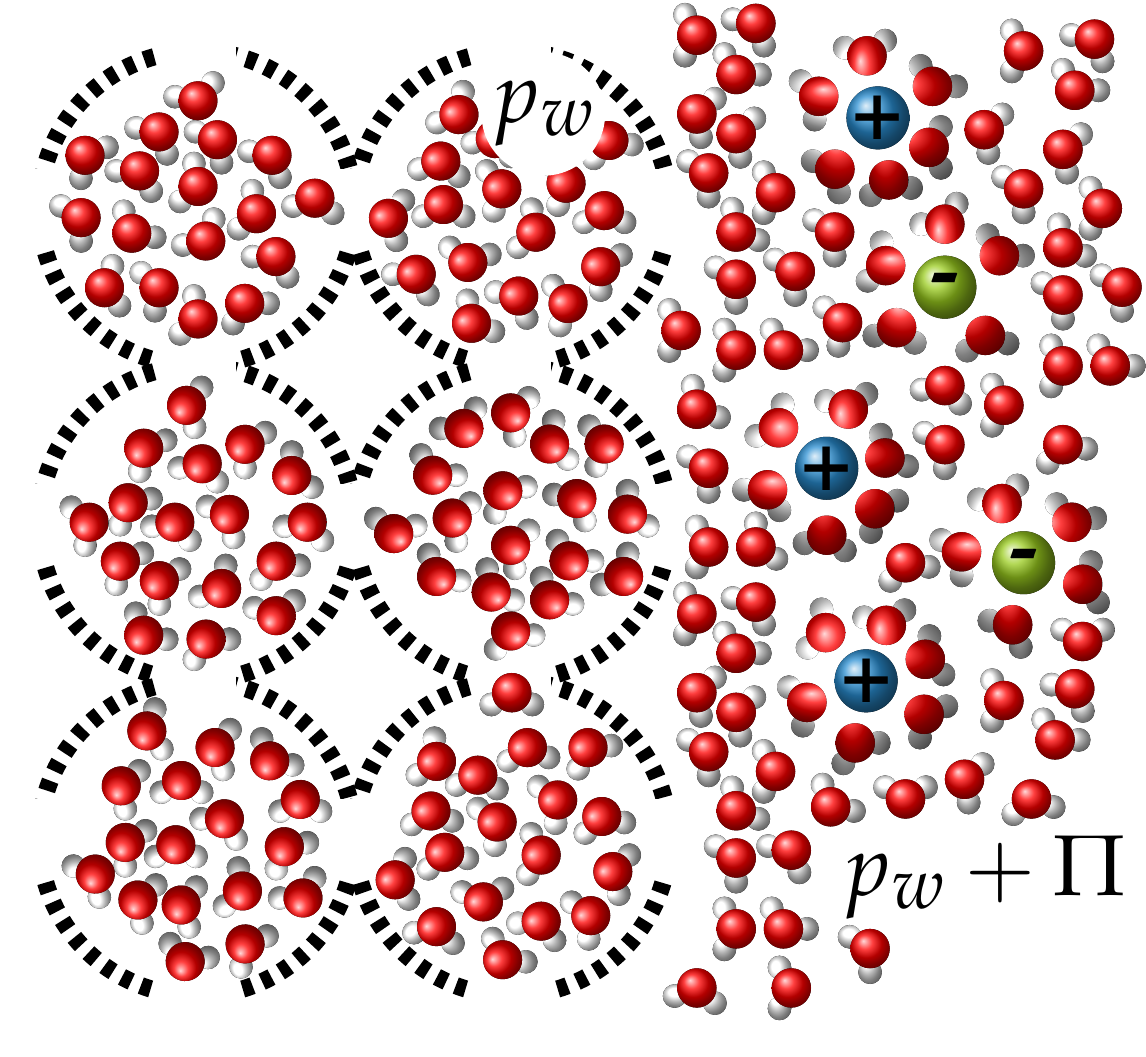}
\caption{Schematic view of the ion exclusion from the ZIF-8 microporous material leading to an osmotic contribution. The jump between the pressure  $p_{w}$ inside the pores of a microporous particle and the pressure $p_w+\Pi$ outside the pores is the osmotic pressure $\Pi$.}
\label{exclusion}
\end{center}
\end{figure}

 This behaviour unravels the mechanism at work in the intrusion-extrusion process of the NaCl solutions. The ZIF-8 behaves as an almost ideal sieve for Na$^+$ and Cl$^-$ ions, as predicted by molecular simulations \cite{Hu2011}. Because of this selectivity, only water penetrates into the material, as sketched in figure~\ref{exclusion}. The pressure difference between the pure water inside the material and the surrounding solution is the osmotic pressure. This pressure difference is constant through the whole process,  therefore the electrolyte cycles are shifted replicas of the pure water cycles, as  only pure water is present into the material. In other words, the pressure shift observed with the electrolyte solution is the additional pressure required to extract pure water from the surrounding electrolyte. 

\begin{figure}[htbp]
\begin{center}
\includegraphics[width=\linewidth]{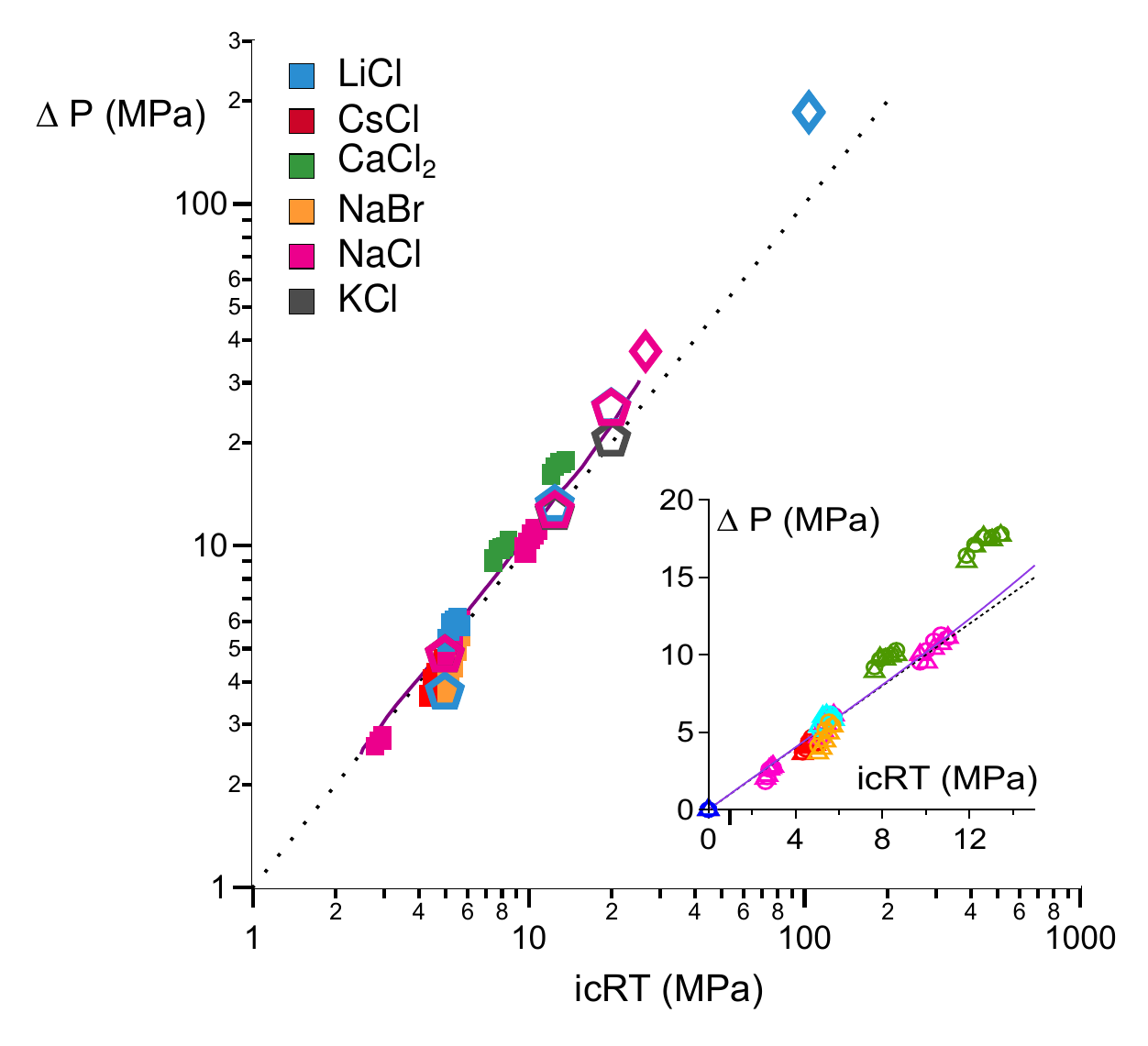}

\caption{Excess  intrusion and extrusion pressures  with respect to pure water measured for the electrolytes mentioned in the legend. The x-scale is the van 't Hoff osmotic pressure $\Pi=icRT$ with $i$ the number of ions per salt molecule and $c$ the salt concentration. The values of our own data ($\blacksquare$) are listed in supplementary Tables 1 to 5 \cite{sup}. 
({\Large $\diamond$}) are data from Tzanis et al \cite{Tzanis2014}, ({\large$\pentagon$}) are data from Ortiz et al \cite{Ortiz2014}. The straight line ($\cdots$) corresponds to the van 't Hoff law of osmotic pressure. The continuous line (\textcolor{Fuchsia}{$\relbar$}) is a  simulation of the osmotic pressure of NaCl solutions from Luo et al  \cite{Luo2010}. The inset is a magnification of our data with ($\Circle$) intrusion pressure and ($\bigtriangleup$) extrusion pressure.}

\label{resume}
\end{center}
\end{figure}

\begin{figure}[htbp]
\begin{center}
\includegraphics[width=0.9\linewidth,]{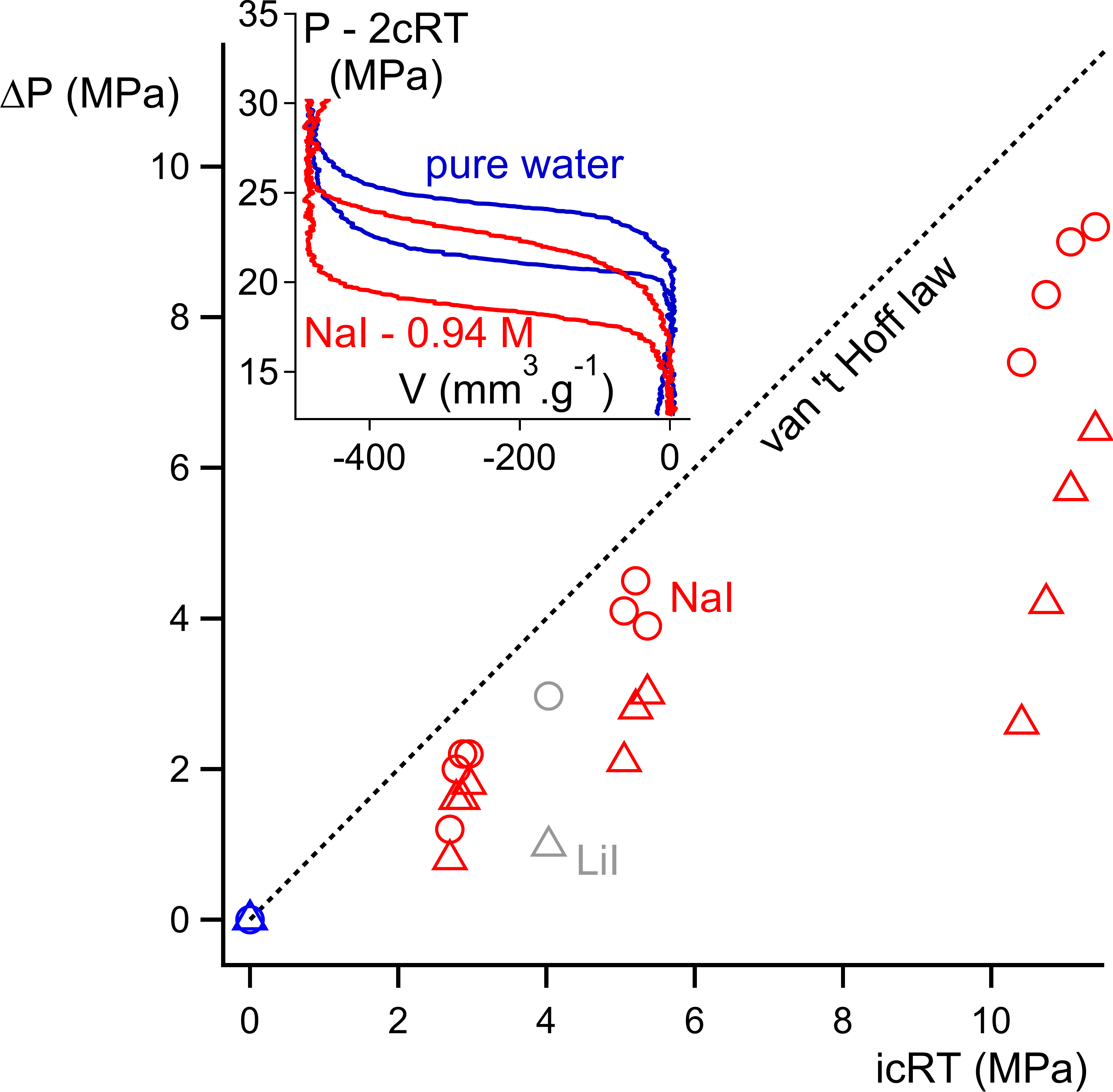}
\caption{Excess  intrusion pressure ($\Circle$) and excess extrusion pressure ($\bigtriangleup$) of NaI and LiI  solutions in ZIF-8 with respect to pure water, plotted as a function of the van~'t Hoff osmotic pressure $icRT$. Inset : difference between the pressure and the van~'t Hoff osmotic pressure of ZIF-8/water (blue) and ZIF-8/NaI - 0.94M (red) at 343 K.}
\label{cycle_NaI}
\end{center}
\end{figure}

We expect the same scenario to hold for all cases in which ions  are perfectly excluded from the microporous material. 
We have studied seven electrolytes: lithium chloride LiCl, lithium iodide LiI, sodium chloride NaCl, sodium bromide NaBr, sodium iodide NaI, calcium chloride CaCl$_2$ and cesium chloride CsCl. 
For each solution, the excess pressure with respect to pure water, $\Delta P= P^{solution}-P^{water}$  was determined both for intrusion and for extrusion at different temperatures \cite{sup}.  
Figure~\ref{resume} summarizes the results with the experimental $\Delta P$'s plotted versus the van 't Hoff pressure $icRT$.  
The figure also includes  data obtained in Silicalite-1 zeolite with NaCl and LiCl \cite{Tzanis2014} and data obtained in ZIF-8 with NaCl, LiCl and KCl \cite{Ortiz2014}. The case of NaI and LiI is discussed later and reported in figure \ref{cycle_NaI}.
As predicted by van't Hoff law, data points obtained for  low concentrations of  LiCl, NaCl, NaBr and CsCl, align along a straight line whose slope is 1.

Measurable deviations from the van 't Hoff law are observed with  saturating  solutions of NaCl and LiCl  in Silicalite-1 zeolites (Tzanis data) as well as with CaCl$_2$-0.5M in ZIF-8 (our data). The deviating pressures are higher than predicted by the van 't Hoff law. We  attribute these deviations to non-ideal effects in the electrolyte solutions. Non-ideal effects are expected when the Coulomb interaction energy between ions becomes comparable to the thermal energy. This occurs for a  typical distance between ions  scaling as $z^{+}z^{-}\lambda_{B}$ with $z^{+}$ and $z^{-}$ the ions valence and $\lambda_{B}$ the Bjerrum length:
\begin{equation}
\lambda_B \sim \frac{e^{2} }{4 \pi \epsilon_0 \epsilon_r k_B T}
\end{equation}
with $\epsilon_0$ the vacuum permitivity, $\epsilon_r$ the water dielectric constant and $k_B$ the Boltzmann constant.
The distance $z^{+}z^{-}\lambda_{B}$ is related to a limit salt concentration $c_{z^{+}z^{-}}$:
\begin{equation}\label{clim}
c_{z^{+}z^{-}}=\frac{1}{i\mathcal{N}_A}\frac{1}{\left(z^{+}z^{-}\lambda_{B}\right)^{3}}=\frac{1}{i\mathcal{N}_A} \left( \frac{4 \pi \epsilon_0 \epsilon_r k_B T}{z^+z^-e^{2}} \right)^3
\end{equation}
and a limit pressure $\Pi_{z^{+}z^{-}}=ic_{z^{+}z^{-}}RT$ above which correlation effects between ions can not be neglected. 
For monovalent ions at the lowest considered temperature $T=303$~K, the limit concentration is around $c_{11}= 1$~M, corresponding to a limit pressure $\Pi_{11}$  of about 5.5 MPa.  
At concentrations  much higher than $c_{11}$  the osmotic pressure is expected to increase more rapidly than the linear van't Hoff prediction. This is the range of concentration above which deviations are experimentally observed  for NaCl and LiCl in figure \ref{resume}. 
 The molecular dynamic simulation of the osmotic pressure of NaCl carried out by Luo et al \cite{Luo2010} indeed meets the data point measured by Tzanis et al  and Ortiz et al for saturating NaCl solutions in zeolithes  \cite{Tzanis2014,Ortiz2014}. Taking into account the trend of the simulation, it is safe to propose that osmotic pressure is also the good candidate to explain  the much higher pressure measured by the same authors for concentrated LiCl solutions (180 MPa of excess pressure at saturation). 

The valence of the ions  has a major impact on $\Pi_{z^{+}z^{-}}$. For a salt with one divalent ion and one monovalent ion such as CaCl$_2$, the limit concentration is $c_{21} = 0.15$~M, and the corresponding pressure is $\Pi_{21}=1.2$~MPa.  Accordingly, the experimental values of $\Delta P$ measured for CaCl$_2$ solutions are larger than the van't Hoff theoretical curve in figure \ref{resume}, and the data points go away from the van't Hoff law as the concentration increases, in a  similar way to what is observed with monovalent ions at higher concentration. 

Remains the particular case of NaI and LiI solutions depicted on figure~\ref{cycle_NaI}. The behaviour of these salts is very different from the previous ones. The cycle of the NaI - 0.94~M solution is compared to one of pure water in the inset of figure~\ref{cycle_NaI}. The electrolyte cycle is not  anymore a simple translation of the pure water cycle. The excess of the  intrusion pressure is much larger than the excess of  the extrusion pressure, and the cycle hysteresis is larger for the electrolyte solution. A similar result is  obtained with all concentrations of NaI and LiI solutions as shown in figure~\ref{cycle_NaI}.
Furthermore the excess pressures measured in the electrolyte cycle with respect to pure water are below the van 't Hoff osmotic pressure.

We attribute this behavior to the partial penetration of ions inside the micropores. As a result the interfacial energies are expected to be different from the one of pure water, and consequently the shape of the cycle is also different. Furthermore the pressure difference between the surrounding electrolyte and the intruding liquid is less than the van 't Hoff osmotic pressure as the intruding liquid  is not pure water. 
The penetration of these ions into the pores is far from intuitive as the radius of the hydrated I$^{-}$, $r_H=3.3$~\AA  \cite{Conway1981},  is twice larger than the radius of the smallest ZIF-8 pores. The partial penetration of  I$^{-}$ ions in ZIF-8 is certainly due to their large polarisability, responsible for their deformability and affinity for hydrophobic surfaces\cite{Huang2007,Jungwirth2006}. 

In any case, the transition between ion intrusion and exclusion is not related to electrostatic screening. Iodide ions enter the pores for small concentrations corresponding to Debye length larger than the pore size while other ions are still excluded from the pore at high concentrations corresponding to Debye length smaller than 0.1 nm.

In summary, our results show that electrolyte solutions can exhibit giant intrusion/extrusion pressures in micropores due to osmotic effects. As ions are repelled from the pores rather than confined in them, the intrusion process is a three-dimensional extraction of pure water from the electrolyte solution.
This osmotic mechanism allows to increase the pressure up to one order of magnitude compared to the intrusion/extrusion pressure obtained with pure water. The intrusion process is associated to a storage of mechanical energy, equal to the product of the macroscopic pressure times the intruded volume\cite{Eroshenko2012}.  
As a result the energy stored in the pores per unit volume can be higher than with ultracapacitors.
Morevover, it was demonstrated that the intrusion/extrusion pressures depend weakly on the rate of the process due the very short time scale of wetting and drying transitions\cite{Guillemot2012}.
Therefore our results pave the way towards the design of lyophobic energy storage systems with targeted power-densities as large as 50 kW per solution liter that is one order of magnitude larger than with ultracapacitors.

From a more fundamental point of view, our results demonstrate experimentally that the concept of osmotic pressure can be extended up to giant values of hundreds of MPa that have never been reached with conventional separation membranes. The selective pulverulent nanomaterial act as a volumic membrane able to sustain huge osmotic pressures because each individual material particle is submitted to a uniform osmotic pressure.
Finally, the forced intrusion of aqueous solution in hydrophobic nanoporous material can be put forward as a way to probe mechanically molecular interactions. This mechanical characterization offers a new way to probe the stability of the hydration layer of ions and the affinity of ions for hydrophobic surfaces as exemplified with the difference of behaviors between iodide ions and other anions.


The authors kindly thank Lyd\'eric Bocquet and Benoit Coasne for fruitful discussions. We thanks also Airbus Defence and Space, the French Direction G\'en\'erale de l'Armement, the AGIR funding from the University Grenoble Alpes and the French National Agency for Research (ANR-14-CE05-0017)  which support our researches.

\nocite{Brunauer1938}

\bibliography{bibliography}

\end{document}